\documentclass{article}
\usepackage{spconf,amsmath,graphicx}
\usepackage{graphicx}
\usepackage{subfigure}
\usepackage{multirow}
\usepackage{color}
\usepackage{stfloats}
\usepackage{hyperref}

\usepackage{adjustbox}
\title{Multi-Field De-interlacing using \\ Deformable  Convolution Residual Blocks and Self-Attention}
\name{Ronglei Ji and~A. Murat Tekalp\thanks{This work was supported in part by TUBITAK 2247-A Award No.~120C156 and KUIS AI Center funded by Turkish Is Bank. A.M. Tekalp also acknowledges support from Turkish Academy of Sciences (TUBA). Ronglei Ji would like to acknowledge a Fung Scholarship.}}
\address{Koc University, Dept. of Electrical and Electronics Engineering, Istanbul, Turkey}
\begin{document}
%
\maketitle
\begin{abstract}
Although deep learning has made significant impact on image/video restoration and super-resolution, learned deinterlacing has so far received less attention in academia or industry. This is despite deinterlacing is well-suited for supervised learning from synthetic data since the degradation model is known and fixed. In this paper, we propose a novel multi-field full frame-rate deinterlacing network, which adapts the~state-of-the-art superresolution approaches to the~deinterlacing task. Our model aligns features from adjacent fields to a reference field (to be deinterlaced) using both deformable convolution residual blocks and self attention. 
Our extensive experimental results demonstrate that the~proposed method provides state-of-the-art deinterlacing results in terms of both numerical and perceptual performance. At~the~time of writing, our model ranks first in the Full FrameRate LeaderBoard at  \url{https://videoprocessing.ai/benchmarks/deinterlacer.html}
\end{abstract}
\vspace{-5pt}
\section{Introduction}
Interlaced scanning was invented to strike a balance between spatial and temporal video resolution in analog TV broadcasting to overcome insufficient bandwidth to transmit  at a high enough frame rate. It corresponds to scanning odd and even fields alternately, which results in doubling the field rate at the expense of vertical spatial resolution without increasing bandwidth, and enhances visual perception of fast motion scenes.

In order to convert interlaced video content to progressive content for displaying interlaced videos on progressive monitors, deinterlacing continues to be a significant problem of interest since there still exists a large amount of interlaced and telecined catalogue content despite the fact that almost all video cameras are progressive nowadays.

The industry employs simple traditional methods for deinterlacing, telecine, and 50 Hz to 60 Hz and vice versa conversion~\cite{tekalp2015}. Many alternative traditional approaches have also been proposed by the academia for intra-field deinterlacing~\cite{yoo2002direction,kim2007novel} and inter-field deinterlacing~\cite{kwon2003deinterlacing},
which, however, introduce either blurring or artifacts such as flicker and jaggedness due to the time difference between adjacent fields.

This paper proposes a novel multi-field deinterlacing network in Section~\ref{method} featuring two stages: a field-alignment~stage consisting of novel deformable convolution residual blocks in parallel with a self-attention module, and a reconstruction module, which generates two 
progressive frames for each pair of input interlaced fields. 
Section~\ref{expr} presents extensive experimental results 
and Section~\ref{conc} concludes the paper. 
\vspace{-10pt}

\section{Related Works and Contributions}
\label{related}

\begin{figure*}[htbp]
\centering
\subfigure[]
{
    \begin{minipage}[h]{0.65\linewidth}
    \flushleft
    \includegraphics[scale=0.45]{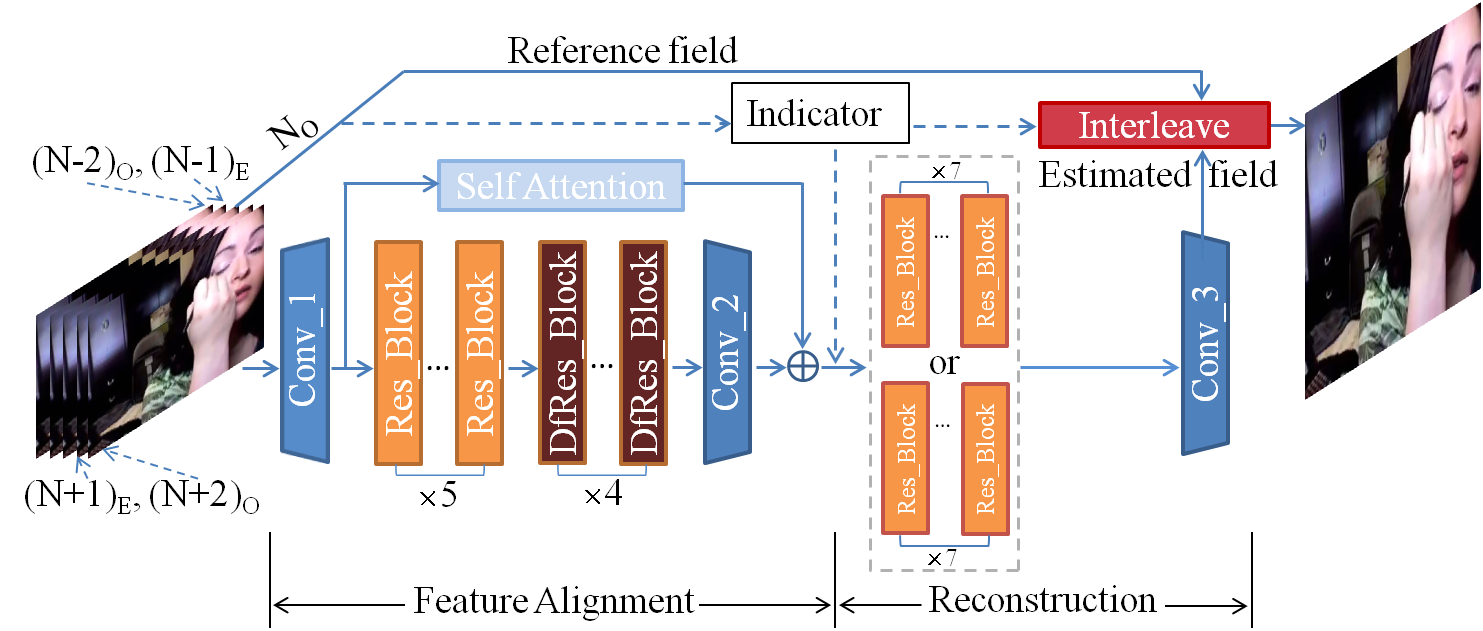}
    \label{fig:DfRes for deinterlacing}
    \vspace{10pt}
    \end{minipage}
}%
\subfigure[]{
    \begin{minipage}[p]{0.35\linewidth}
    \includegraphics[width=4.5cm]{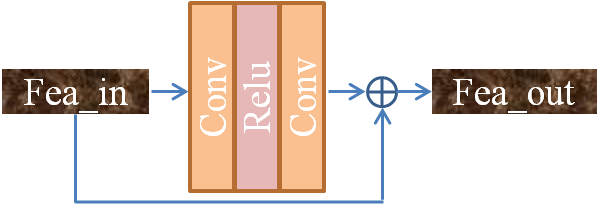} \vspace{2pt}
    \label{fig:Res block}
    \vspace{2pt}
    \hspace{10mm}
   \\
    \includegraphics[width=6.1cm]{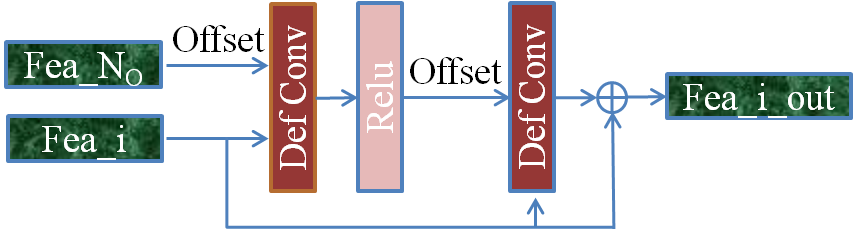} \vspace{2pt}
    \label{fig:DfRes block}
    \\
    \includegraphics[width=6.1cm]{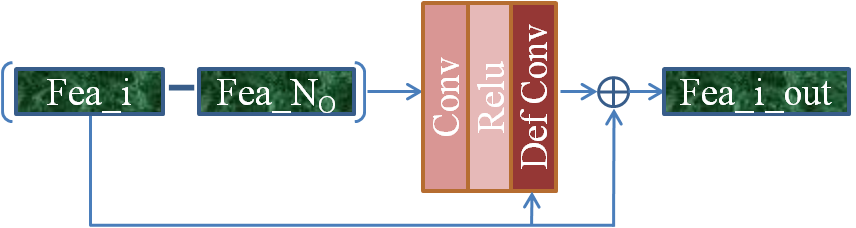} \vspace{2pt}
    \label{fig:DfRes block}
    
    \end{minipage}
}
\vspace{-10pt}
\caption{(a) The proposed deinterlacing network with five input fields $(N-2)_O, (N-1)_E, N_O, (N+1)_E, (N+2)_O$, where the reference field for alignment is~$N_O$, the field to be estimated is $N_E$.
(b) Top, block diagram of a standard residual block depicted by orange boxes~in~(a). Middle, block diagram of a DfRes block shown by the brown box in (a).
Bottom, block diagram of a differential DfRes ($\Delta$DfRes) block. Fea\_i represents each input field. Fea\_i\_out is the corresponding output feature after one DfRes ($\Delta$DfRes) block and it will replace Fea\_i to be aligned through next DfRes ($\Delta$DfRes) block.} 
\label{fig} 
\end{figure*}

\textbf{Related Works.} There are only few prior works that have attempted deinterlacing using deep learning \cite{zhu2017real,akyuz2020deep,bernasconi2020deep}. 
Yet, \cite{zhu2017real,akyuz2020deep} consider only intra-frame deinterlacing and \cite{bernasconi2020deep} does not employ feature-level alignment of supporting fields; hence, their results are suboptimal in the presence of large motion. As deformable convolution has been adopted in many video processing tasks~\cite{dai2017deformable,tian2020tdan,wang2019edvr,yilmaz2021dfpn},
we presented one of the first works on feature-level field alignment using deformable convolutions specialized for the deinterlacing task~\cite{ji2021learned}. This paper is an improved edition of~\cite{ji2021learned}, and achieves comparable performance to recent papers~\cite{zhao2021multi,liu2021spatial} that also adopt deformable convolution for feature-level alignment for the deinterlacing. However, different from \cite{ji2021learned}, with the excellent results in computer vision applications achieved by self attention (SA)~\cite{vaswani2017attention},
this paper proposes to combine deformable convolution residual (DfRes) block and self-attention block in parallel to align field feature for full-frame rate deinterlacing.

\noindent\textbf{Contributions.} This paper advances the state-of-the-art in learned deinterlacing as follows:
\par 1) Different from our early work~\cite{ji2021learned}, we employ a self-attention module in parallel to the DfRes blocks and show that this new architecture provides improved performance by enabling better multi-frame feature alignment and fusion.
\par 2) We propose to utilize separate reconstruction module to restore even and odd fields separately, with which the overall performance of all methods presented here are significantly improved over those in~\cite{ji2021learned}.
\par 3) 
We provide ablation and generalization results to show the effectiveness of proposed model over competing learned
deinterlacing methods.
\vspace{-5pt}

\section{Proposed Deinterlacing Architecture}
\label{method}
\vspace{-5pt}
\vspace{-5pt}
\subsection{Overview
of the Proposed Architecture}
\label{netarch}
\vspace{-5pt}

The proposed overall architecture is depicted in Fig.~\ref{fig}~(a).
The input to the field feature alignment stage consists of an odd number of fields centered about the reference field.
We use an indicator bit, which is 1 when the reference field is an even field; or 0 when the reference field is an odd field, when forming the output progressive frames.
\par Each data field is first mapped into 64 feature channels via the convolution layer Conv\_1 and next processed by five regular residual blocks. These features are then fed into four separate alignment modules to align the features of each supporting field to those of the reference field using deformable convolution residual blocks (DfRes\_Block $\times 4$). The aligned features are then stacked and they are fused and reduced to 64 channels through Conv\_2.

In addition, we employ an SA module in parallel to the DfRes blocks to enhance the performance of feature registration. The 64-channel feature tensors processed by the SA module and DfRes module in parallel are added together to form the final 64-channel aligned feature tensor. Experiments demonstrate the effectiveness of the proposed strategy.

The 64-channel aligned feature tensor is input to the reconstruction stage, where it is processed by two branches of 7~regular residual blocks each. The last Conv\_3 layer maps 64~channels to 3 channels to generate the estimated field.

\begin{figure}[t!]
\centering
\includegraphics[scale=0.35]{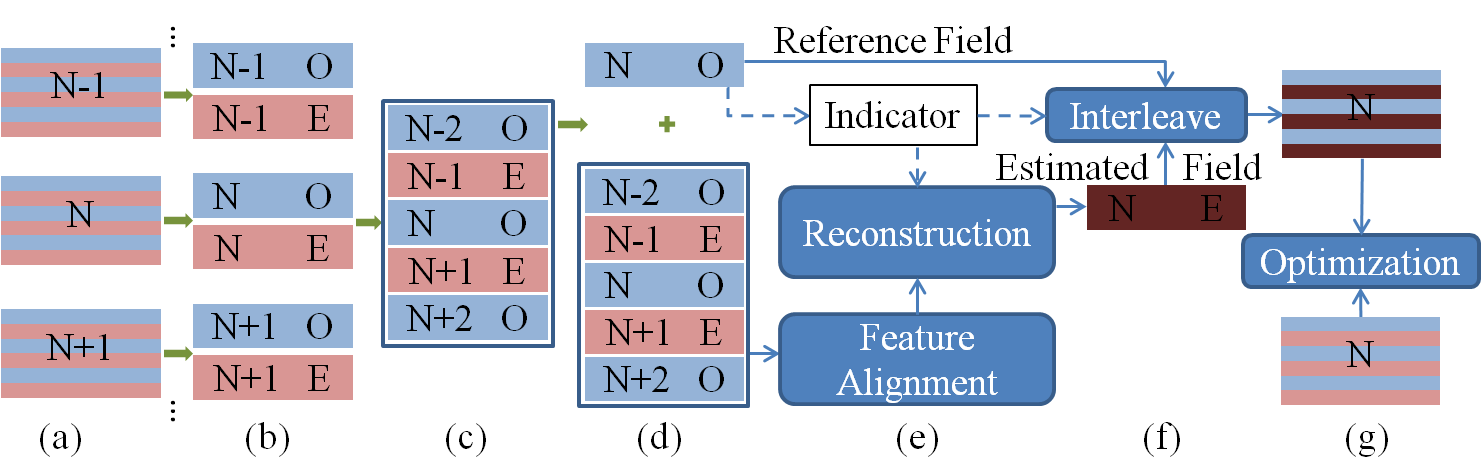}
\vspace{-5pt}
\caption{Overview of data processing during training. (a)-(b) synthetic interlaced videos are generated by extracting odd and even fields of video frames, where $N_O$ and $N_E$ represent odd and even fields of frame $N$. (c) input fields. (d)-(g) deinterlacing process.}
\label{fig:dataprocessing}
\end{figure}

Finally, we interleave the available reference field and estimated opposite parity field using the indicator bit to form progressive output frames as shown in Fig.~\ref{fig:dataprocessing}~(d)~(f)~(g). If~the indicator field is 0 (1), we place the estimated field in the even (odd) lines of the output progressive frame.
\vspace{-5pt}

\subsection{Field Alignment using Deformable Convolution Blocks}
\label{proposed}
\vspace{-5pt}


Deformable convolution residual (DfRes) block~\cite{ji2021learned}, shown in Fig.~\ref{fig}~(b) middle diagram
employs two deformable convolution layers instead of regular convolutions in order to align features of supporting fields with those of the reference field. 

$\Delta$DfRes Block~\cite{ji2021learned} 
has similar performance but less parameters compared to DfRes block,
and it is depicted in Fig.~\ref{fig}~(b) bottom diagram.


\vspace{-5pt}
\subsection{Field Alignment using Efficient Self-Attention}
\label{selfattention}

\begin{figure}[t!]
\centering
\subfigure[]
{
    \begin{minipage}[t]{0.5\linewidth}
    \centering
    \includegraphics[scale=0.27]{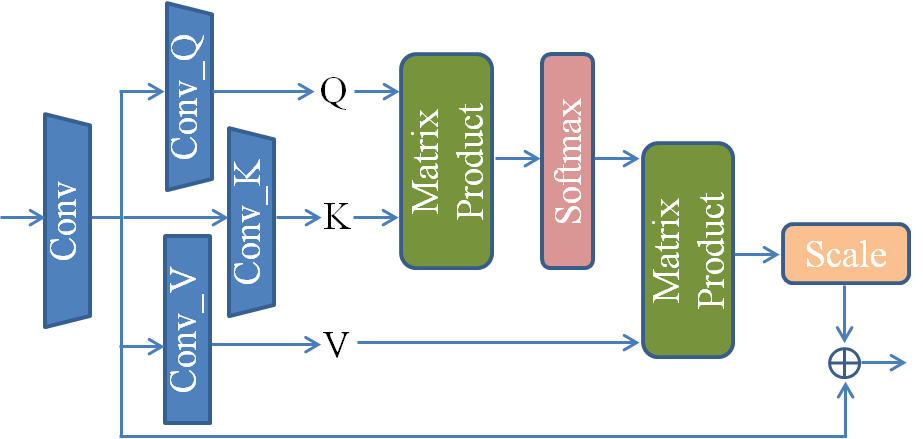} \vspace{-4pt}
    \label{fig:SA}
    \end{minipage}
}%
\subfigure[]
{
   \begin{minipage}[t]{0.5\linewidth}
    \center
    \includegraphics[scale=0.27]{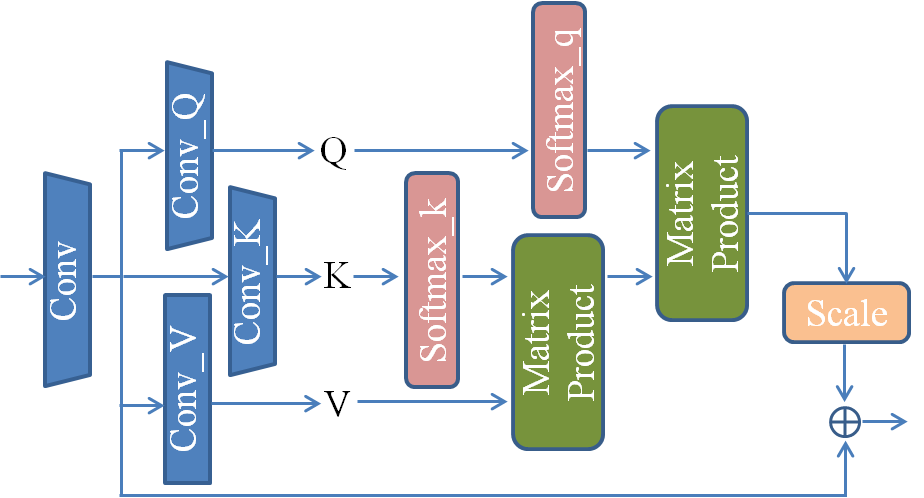} \vspace{-4pt}
    \label{fig:ESA}
    \end{minipage}
} 
\caption{(a) Block diagram of the self-attention module, where Softmax is defined in Equation~\ref{eq:softmax}. (b) Block diagram of the efficient self-attention module, where Softmax\_k and Softmax\_q are defined in Equation~\ref{eq:twosoftmax}.} 
\label{fig:selfattn} 
\end{figure}
\vspace{-5pt}

The SA module shown in dashed light blue box in Fig.~\ref{fig}~(a) and depicted in detail in Fig.~\ref{fig:selfattn}~(a), is parallel to feature extraction Res\_Block~$\times 5$ module and feature alignment DfRes module. The data fields with $n$ pixels in each field are first mapped to 64 channel feature space via regular convolution operations (Conv\_1 in Fig.~\ref{fig}~(a) and Conv in Fig.~\ref{fig:selfattn}~(a)). Then, queries (Q $\in {R}^{8 \times n} $),  keys (K $\in {R}^{8 \times n} $) and values (V $\in {R}^{64 \times n} $) are computed by regular convolution operations Conv\_Q, Conv\_K and Con\_V separately, where the output feature channels of Conv\_Q and Conv\_K are $\frac{1}{8}$ times of those input feature channels as shown with blue trapezoids in Fig.~\ref{fig:selfattn}~(a), and the input and output channels of Con\_V are the same as shown with blue rectangle in Fig.~\ref{fig:selfattn}~(a). After obtaining Q, K and V matrix, we multiply Q and K by transposing Q and feed the result into softmax operation $\rho(Q^{T} \times K)$ for normalization. Finally, we multiply V and the transpose of the normalized softmax output to gain an initial output of Self Attention Module. The final output of Self Attention Module is the original 64 channel feature resulted from Conv in Fig.~\ref{fig:selfattn}~(a) plus the scaled initial output as shown in plus sign in Fig.~\ref{fig:selfattn}~(a). Hence, the Self Attention Module operation can be expressed in the following equation:
\vspace{-5pt}
\begin{equation}
\begin{split}
SA(Q,K,V) = V \times (\rho(Q^{T} \times K))^{T} \cdot Scale,
\label{eq:SA}
\end{split}
\end{equation}
where T denotes matrix transpose operation and $\times$ matrix product. Scale is a learnable factor for scaling initial output. We use softmax function for normalization:
\vspace{-5pt}
\begin{equation}
\begin{split}
Softmax:~\rho(Y) = \sigma_{row}(Y),
\label{eq:softmax}
\end{split}
\end{equation}
where $\sigma_{row} $ applys softmax rowwisely.
\par In dealing with the huge computation cost of self attention, we utilize the following approximate equation in test~\cite{shen2021efficient}:
\vspace{-5pt}
\begin{equation}
\begin{split}
\vspace{-5pt}
ESA(Q,K,V) = V \times \rho_k(K^{T}) \times \rho_q(Q) \cdot Scale,
\vspace{-5pt}
\label{eq:esa}
\end{split}
\end{equation}
where $\rho_k$ and $\rho_q$ are softmax functions utilized to normalize K and Q features. They are defined as:
\vspace{-5pt}
\begin{equation}
\begin{aligned}
\begin{split}
Softmax\_k:~&\rho_k(Y) = \sigma_{col}(Y), \\
Softmax\_q:~&\rho_q(Y) = \sigma_{row}(Y),
\label{eq:twosoftmax}
\end{split}
\end{aligned}
\end{equation}
where $\sigma_{col} $ and $\sigma_{row} $ apply softmax columnwisely and rowwisely, respectively.
\par Efficient Self Attention Module as shown in Fig.~\ref{fig:selfattn}~(b) reduces test time to the level of models without self attention module with only about 0.02 dB PSNR loss, which generates a close approximation due to the non-linearity of softmax function. We call the DfRes module based deinterlacing network with this parallel Self Attention Module DfRes\_SA.

\subsection{Frame Reconstruction Module}
\label{sep}
\vspace{-5pt}
We propose to employ two same reconstruction blocks in parallel (i.e., Res\_Block $\times 7$, right two orange box in Fig.~\ref{fig}~(a), and details are in Fig.~\ref{fig}~(c)) to reconstruct even and odd fields separately in each data batch, which can restore fields more pertinently and directionally and thus produce better performance than the model with only one reconstruction block. We call these two parallel reconstruction blocks Separate Reconstruction Module. We utilize this proposed module in all three proposed DfRes designs, which are DfRes, $\Delta$DfRes and DfRes\_SA, and In Section~\ref{expr}, we will compare deinterlacing results from these three designs.

\vspace{-5pt}
\section{Experimental Evaluation} 
\label{expr}

\subsection{Experimental Settings}
\label{setting}
\vspace{-5pt}
\textbf{Dataset.} We used UCF101~\cite{soomro2012ucf101}, REDS~\cite{wang2019edvr,nah2019ntire} and Vimeo-90K~\cite{xue2019video} datasets with the same separation in~\cite{ji2021learned} to demonstrate the effectiveness of our method and compare it with other methods. 
We also used Vid4~\cite{sajjadi2018frame} dataset for testing. 


\noindent\textbf{Evaluation Methods.} Peak signal-to-noise ratio (PSNR) and structural similarity index (SSIM) are used to quantitatively evaluate results and zooming in on details for subjective evaluation of the results.

\noindent\textbf{Training Procedure.} All the training settings are the same with~\cite{ji2021learned} except that we train our model for 150,000 iterations using Pytorch over tesla\_v100 and that in the loss function in~\cite{ji2021learned} we replace $SSIM(Y^{pred},Y^{GT})$ with $0.1 \cdot Cb(Y^{pred},Y^{GT})$, where $Cb(\cdot)$ is Charbonnier Loss.

\begin{table}[t]
\vspace{0in}
\setlength{\abovecaptionskip}{0in}
\setlength{\belowcaptionskip}{0in} 
\caption{Comparison of PSNR (top) and SSIM (bottom) for different methods on multiple datasets. \textcolor{red}{The best} and \textcolor{blue}{the second best} are marked with red and blue color respectively. \vspace{-6pt} }
\begin{center}
\begin{tabular}{ c||c|c|c|c|c }
\hline
 Datasets
 & \textbf{DfRes\_SA} & \textbf{DfRes}
 & EDVR
 & TDAN & DUF \\\hline \hline
\multirow{2}*{UCF101} 
 & \textcolor{red}{51.92}
 & \textcolor{blue}{51.48}
 & 51.23
 & 47.74
 & 49.75
 \\ 
 & \textcolor{red}{0.99931}
 & \textcolor{blue}{0.9993}
 & 0.9992
 & 0.998
 & 0.999
 \\  \hline
 
 \multirow{2}*{Vimeo}
 & \textcolor{red}{44.35}
 & \textcolor{blue}{43.45}
 & 42.15
 & 40.73
 & 40.66
 \\ 
 & \textcolor{red}{0.9949}
 & \textcolor{blue}{0.994}
 & 0.991
 & 0.9889
 & 0.989
 
 \\  \hline
 \multirow{2}*{REDS} 
 & \textcolor{red}{37.00}
 & \textcolor{blue}{36.51}
 & 33.20
 & 30.09
 & 31.36
 \\
 & \textcolor{red}{0.984}
 & \textcolor{blue}{0.982}
 & 0.964
 & 0.944
 & 0.955
 \\  \hline
\end{tabular}
\label{on3dataset}
\end{center}  \vspace{-16pt}
\end{table}

\begin{table}[t]
\vspace{0in}
\setlength{\abovecaptionskip}{0in}
\setlength{\belowcaptionskip}{0in} 
\caption{Comparison of regular offset estimation and proposed DfRes offset estimation on UCF101 dataset. \vspace{-3pt} }
\begin{center}
\begin{tabular}{ c||c|c }
\hline
 Methods & Regular + SA & DfRes\_SA \\\hline\hline
 PSNR
 & 51.89
 & 51.92
 \\ \hline
 SSIM
 & 0.99929
 & 0.99931
 \\  \hline 
  
\end{tabular}
\label{difoffset}
\end{center}  \vspace{-20pt}
\end{table}

\begin{table}[t]
\vspace{0in}
\setlength{\abovecaptionskip}{0in}
\setlength{\belowcaptionskip}{0in} 
\caption{Ablation study of proposed method. \vspace{-3pt} }
\begin{center}
\begin{tabular}{ c||c|c|c }
\hline
 Modules & DfRes+SA & only DfRes & only SA \\
 \hline \hline
 PSNR
 & 51.92
 & 51.48
 & 50.23
 \\ \hline
 SSIM
 & 0.99931
 & 0.99925
 & 0.99907
  \\  \hline
\end{tabular}
\label{ablation}
\end{center}  \vspace{-6pt}
\end{table}

\vspace{-10pt}

 
 
  

\subsection{Comparison with the State-of-the-Art methods}
\label{comp}
We compare our proposed DfRes, $\Delta$DfRes and DfRes\_SA deinterlacing methods with three state-of-the-art SR methods: EDVR \cite{wang2019edvr}, TDAN \cite{tian2020tdan} and DUF \cite{jo2018deep}. 
We plug modules of these three comparison networks into our deinterlacing architecture by replacing modules in Fig.~\ref{fig:dataprocessing} (e) with corresponding modules of the comparison networks. We do not modify these comparison methods except that we change the upsampling step to direct combination of reference field and estimated field specifically for deinterlacing as shown in Fig.~\ref{fig:dataprocessing} (f)-(g).

\par The quantitative comparison results in Table \ref{on3dataset} show that on three datasets, our proposed DfRes
and DfRes\_SA methods outperform all other methods in terms of PSNR and SSIM. 

\par
\begin{figure}[t!]
\subfigure[DfRes\_SA]
{   \begin{minipage}[t]{0.21\linewidth}
   \includegraphics[scale=1.9]{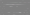}\\
   \vspace{-5mm}
    \label{fig:f0}
    \end{minipage}
}\hspace{0.01mm}
\subfigure[DfRes]
{   \begin{minipage}[t]{0.21\linewidth}
   \includegraphics[scale=1.9]{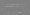}\\
   \vspace{-5mm}
    \label{fig:f0}
    \end{minipage}
}\hspace{0.01mm}
\subfigure[$\Delta$DfRes]
{   \begin{minipage}[t]{0.21\linewidth}
   \includegraphics[scale=1.9]{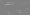}\\
   \vspace{-5mm}
    \label{fig:f0}
    \end{minipage}
}\hspace{0.01mm}
\subfigure[EDVR]
{   \begin{minipage}[t]{0.22\linewidth}
    \includegraphics[scale=1.9]{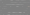}
    \vspace{-5mm}
    \label{fig:f0}
    \end{minipage}
}
\\
\subfigure[TDAN]
{   \begin{minipage}[t]{0.21\linewidth}
    \includegraphics[scale=1.9]{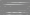}
    \vspace{-5mm}
    \label{fig:f0}
    \end{minipage}
}\hspace{0.01mm}
\subfigure[DUF]
{   \begin{minipage}[t]{0.21\linewidth}
    \centering
    \includegraphics[scale=1.9]{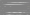}
    \vspace{-5mm}
    \label{fig:f0}
    \end{minipage}
}\hspace{0.01mm}
\subfigure[\cite{zhu2017real}]
{   \begin{minipage}[t]{0.21\linewidth}
    \includegraphics[scale=1.9]{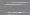}
    \vspace{-5mm}
    \label{fig:f0}
    \end{minipage}
}\hspace{0.01mm}
\subfigure[EDVR\_woTSA]
{   \begin{minipage}[t]{0.24\linewidth}
   \includegraphics[scale=1.9]{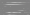}\\
   \vspace{-5mm}
    \label{fig:f0}
    \end{minipage}
}\hspace{0.01mm}
\centering
\subfigure[]
{
    \begin{minipage}[t]{0.23\linewidth}
    \vspace*{0.1in}
    \centering
    \includegraphics[scale=0.5]{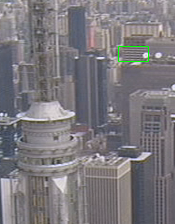}
    \label{fig:h0}
    \end{minipage}
}
\subfigure[]
{   \begin{minipage}[t]{0.75\linewidth}
    \vspace{0.1in}
    \centering
    \includegraphics[scale=1.7]{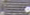}\vspace{1pt}
    \includegraphics[scale=1.7]{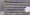}\vspace{1pt}
    \includegraphics[scale=1.7]{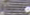}\vspace{1pt}
    \hspace{10mm}
    \\
    \includegraphics[scale=1.7]{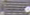}\vspace{1pt}
    \includegraphics[scale=1.7]{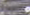}\vspace{1pt}
    \includegraphics[scale=1.7]{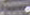}\vspace{1pt}
    \hspace{10mm}
    \\
    \includegraphics[scale=1.7]{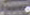}\vspace{4pt}
    \includegraphics[scale=1.7]{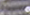}\vspace{4pt}
    \includegraphics[scale=1.7]{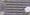}\vspace{4pt}
    \hspace{10mm}
    \\
    \end{minipage}
}
\vspace{-10pt}
\caption{Visual evaluation of a deinterlaced frame from Vid4 dataset. (a)-(h) Differences between the actual and reconstructed progressive frames. (i) Progressive ground truth frame. (j)~Zooming in the green box. From upper left to lower right: DfRes\_SA (PSNR: 37.29), DfRes (PSNR: 37.15), $\Delta$DfRes (PSNR: 37.25), EDVR (PSNR: 36.93), EDVR\_woTSA (PSNR: 36.21), TDAN (PSNR: 35.41), DUF (PSNR: 36.08), \cite{zhu2017real} (PSNR: 29.31), and ground truth. } 
\label{resultsonVid} 
\end{figure}

 Visual comparisons in Fig.~\ref{resultsonVid},
which include EDVR method without attention module (EDVR\_woTSA), show that our proposed DfRes\_SA, DfRes and $\Delta$DfRes methods produce sharper frames and EDVR method plugged into our deinterlacing architecture can also yield comparable visual effect.

\par In \cite{dai2017deformable,tian2020tdan,wang2019edvr},
offsets are estimated by regular convolution operation on the concatenation of neighboring field and reference field. In Table~\ref{difoffset}, we demonstrate the improvement of proposed DfRes offset estimation over regular one. 

\par Ablation study in Table~\ref{ablation} shows that Deformable Convolution and Self Attention module can complement each other for feature alignment to improve performance.

\begin{table*}[b!]
\setlength{\abovecaptionskip}{0pt}
\setlength{\belowcaptionskip}{0pt}
\caption{Generalization of the proposed DfRes, $\Delta$DfRes and DfRes\_SA methods trained and tested on different datasets.} \vspace{0pt}
\begin{center}
\begin{tabular}{ c||c c|c|c||c|c|c||c|c|c }
\hline
  {\multirow{2}{*}{Train on}}
   & 
   & \multicolumn{3}{c||}{\textbf{DfRes\_SA:}
   Test on } 
   & \multicolumn{3}{c||}{\textbf{DfRes:} Test on } 
   & \multicolumn{3}{c}{\textbf{$\Delta$DfRes:} Test on } \\ \cline{3-11} 
   & 
   & UCF101 & Vimeo & REDS4 
   & UCF101 & Vimeo & REDS4
   & UCF101 & Vimeo & REDS4 \\ \hline\hline
\multirow{2}{*}{UCF101} 
 & PSNR
 & 51.92
 & 40.78
 & 34.86
 & 51.48
 & 40.59
 & 34.40
 & 51.46
 & 38.99
 & 34.39
  \\  \cline{2-2}
 & SSIM
 & 0.999
 & 0.990
 & 0.975
 & 0.999
 & 0.990
 & 0.972
 & 0.999
 & 0.984
 & 0.972
 \\ \hline
\multirow{2}{*}{Vimeo} & PSNR
 & 46.63
 & 44.35
 & 36.00
 & 46.06
 & 43.45
 & 35.20
 & 46.11
 & 43.71
 & 35.39
 \\  \cline{2-2}
 & SSIM
 & 0.998
 & 0.995
 & 0.980
 & 0.998
 & 0.994
 & 0.976
 & 0.998
 & 0.994
 & 0.977
 \\ \hline
\multirow{2}{*}{REDS} & PSNR
 & 38.04
 & 42.39
 & 37.00
 & 37.26
 & 42.04
 & 36.51
 & 37.78
 & 41.82
 & 36.49
 \\  \cline{2-2}
 & SSIM
 & 0.9919
 & 0.9933
 & 0.9835
 & 0.9917
 & 0.9928
 & 0.9817
 & 0.9915
 & 0.9924
 & 0.9816
\\ \hline
 \end{tabular}
\label{generalization}
\end{center}  
\end{table*}


Visual comparisons shown in Fig. \ref{resultsonVid} illustrate that our proposed methods produce sharper results than \cite{zhu2017real}, which is also neural network based full frame rate deinterlacing methods.

More quantitative and qualitative comparison results concerning proposed methods for same category deinterlacing methods can be found on MSU benchmark website\cite{MSU}.
\vspace{-5pt}

\subsection{Evaluation of Generalization Performance}
\label{gen}
\par Table~\ref{generalization} shows that our method in general generalizes well but the performance varies by the amount of motion in the training and test sets. Extensive experimental results reveal that both training and test dataset affect the performance of deinterlacing methods.


\section{Conclusion} \vspace{-5pt}
\label{conc}
\par We introduce
a new network architecture for video deinterlacing with feature alignment by combining deformable convolution residual block and self attention, which can incorporate state of the art superresolution approaches for deinterlacing task. Experimental results demonstrate that the proposed network yields state-of-the-art performance and can be adopted to effectively deinterlace videos.
\clearpage

\vfill\pagebreak


\bibliographystyle{IEEEbib}
\bibliography{referenceicip2022}

\end{document}